\documentclass[sigconf]{acmart}
\usepackage{times}  
\usepackage{helvet}  
\usepackage{courier}  
\usepackage{algorithm}
\usepackage{algorithmic}
\usepackage{titletoc}

\usepackage{amsmath,amsfonts,bm}

\def\eqref#1{equation~\ref{#1}}

\def\1{\bm{1}}

\DeclareMathAlphabet{\mathsfit}{\encodingdefault}{\sfdefault}{m}{sl}
\SetMathAlphabet{\mathsfit}{bold}{\encodingdefault}{\sfdefault}{bx}{n}

\newcommand{\sigmoid}{\sigma}

\usepackage{booktabs}
\usepackage{newfloat}
\usepackage{listings}
\usepackage{makecell}
\usepackage{threeparttable}
\usepackage{pdfpages}
\usepackage{import}
\usepackage{paralist}
\usepackage{subcaption}
\usepackage[titletoc]{appendix}
\usepackage{multirow}

 \usepackage{enumitem}
\AtBeginDocument{%
  \providecommand\BibTeX{{%
    \normalfont B\kern-0.5em{\scshape i\kern-0.25em b}\kern-0.8em\TeX}}}

\copyrightyear{2022}
\acmYear{2022}
\setcopyright{acmcopyright}\acmConference[WWW '22]{Proceedings of the ACM Web
Conference 2022}{April 25--29, 2022}{Virtual Event, Lyon, France}
\acmBooktitle{Proceedings of the ACM Web Conference 2022 (WWW '22), April
25--29, 2022, Virtual Event, Lyon, France}
\acmPrice{15.00}
\acmDOI{10.1145/3485447.3512088}
\acmISBN{978-1-4503-9096-5/22/04}

\begin{document}

\title{Multiple Choice Questions based Multi-Interest Policy Learning for Conversational Recommendation}



\author{Yiming Zhang}
\authornote{Both authors contributed equally to this research.}
\affiliation{%
  \institution{Tongji University}
  \country{China}}
   
\email{2030796@tongji.edu.cn}

\author{Lingfei Wu}
\authornotemark[1]
\affiliation{%
  \institution{JD.COM Silicon Valley Research Center}
  \country{USA}}
\email{lwu@email.wm.edu}

\author{Qi Shen}
\affiliation{%
  \institution{Tongji University}
  \country{China}}
   
\email{1653282@tongji.edu.cn}

\author{Yitong Pang}
\affiliation{%
  \institution{Tongji University}
  \country{China}}
   
\email{1930796@tongji.edu.cn}

\author{Zhihua Wei}
\authornote{Corresponding author.}
\affiliation{%
  \institution{Tongji University}
   \country{China}}
  
\email{zhihua_wei@tongji.edu.cn}

\author{Fangli Xu}
\affiliation{%
  \institution{Squirrel AI Learning}
  \country{USA}}
\email{fxu02@email.wm.edu}

\author{Bo Long}
\affiliation{
  \institution{JD.COM}
  \country{China}}
   
\email{bo.long@jd.com}

\author{Jian Pei}
\affiliation{
  \institution{Simon Fraser University}
  \country{Canada}}
\email{jpei@cs.sfu.ca}



\renewcommand{\shortauthors}{Zhang, et al.}

\begin{abstract}
Conversational recommendation system (CRS) is able to obtain fine-grained and dynamic user preferences based on interactive dialogue. Previous CRS assumes that the user has a clear target item, which often deviates from the real scenario, that is for many users who resort to CRS, they might not have a clear idea about what they really like.
Specifically, the user may have a clear single preference for some attribute types (e.g. brand) of items, while for other attribute types (e.g. color), the user may have multiple preferences or even no clear preferences, which leads to multiple acceptable attribute instances (e.g. black and red) of one attribute type. Therefore, the users could show their preferences over items under multiple combinations of attribute instances rather than a single item with unique combination of all attribute instances.
As a result, we first propose a more realistic conversational recommendation learning setting, namely Multi-Interest Multi-round Conversational Recommendation (MIMCR), where users may have multiple interests in attribute instance combinations and accept multiple items with partially overlapped combinations of attribute instances. 
To effectively cope with the new CRS learning setting, in this paper, we propose a novel learning framework, namely Multiple Choice questions based Multi-Interest Policy Learning (MCMIPL). 
In order to obtain user preferences more efficiently, the agent generates multiple choice questions rather than binary yes/no ones on specific attribute instance. Furthermore, we propose a union set strategy to select candidate items instead of existing intersection set strategy in order to overcome over-filtering items during the conversation. 
Finally, we design a Multi-Interest Policy Learning (MIPL) module, which utilizes captured multiple interests of the user to decide next action, either asking attribute instances or recommending items.
Extensive experimental results on four datasets demonstrate the superiority of our method for the proposed MIMCR setting. The implementation of our proposed models is publicly available at \url{https://github.com/ZYM6-6/MCMIPL}.

\end{abstract}

\ccsdesc[500]{Information systems~Users and interactive retrieval}
\ccsdesc[500]{Information systems~Recommender systems}

\keywords{Conversational Recommendation, Reinforcement Learning, Graph Representation Learning}

\maketitle

\section{Introduction}
Conversational recommendation system (CRS) aims to obtain fine-grained and dynamic user preferences and make successful recommendations through conversations with users \cite{EAR,zhang2018towards}. In each conversation turn, CRS can select different actions \cite{lei2020conversational} based on user feedback, either asking attributes or recommending items. Since it is able to explicitly obtain user preferences and has the advantage of conducting explainable recommendation, CRS has become one of the hot topics in current research.

Various methods  \cite{Abs_Greedy,li2021seamlessly,lee2019melu,zou2020neural} have been proposed to improve the performance of CRS based on different problem settings.
In this work, we focus on the multi-round conversational recommendation (MCR) setting \cite{EAR,SCPR,Unicorn}, which is the most realistic CRS setting so far. The system focuses on whether asking attributes or recommending items in each turn, and adjusts actions flexibly via user feedback to make successful recommendations with fewer turns.


Despite the success of MCR in recent years, the assumption of the existing MCR \cite{EAR}, that the user preserves clear preferences towards all the  attributes  and  items, may often deviate from the real scenario.
For the user who resorts to CRS, he might not have a clear idea about what he really likes. Specifically, the user may have a clear single preference for some \textbf{attribute types} (e.g., color) of items, while for other attribute types (e.g., brand), the user might have multiple preferences or even no clear preferences. With the guidance of CRS, he may accept multiple \textbf{attribute instances} (e.g., red and black) of one attribute type. In addition, different combinations of these attribute instances are generally associated with different items. Therefore, the user could show his preferences over items under multiple combinations of attribute instances rather than a single item with unique combination of all attribute instances.



\begin{figure}[t]
    \centering
    \includegraphics[width=0.49\textwidth]{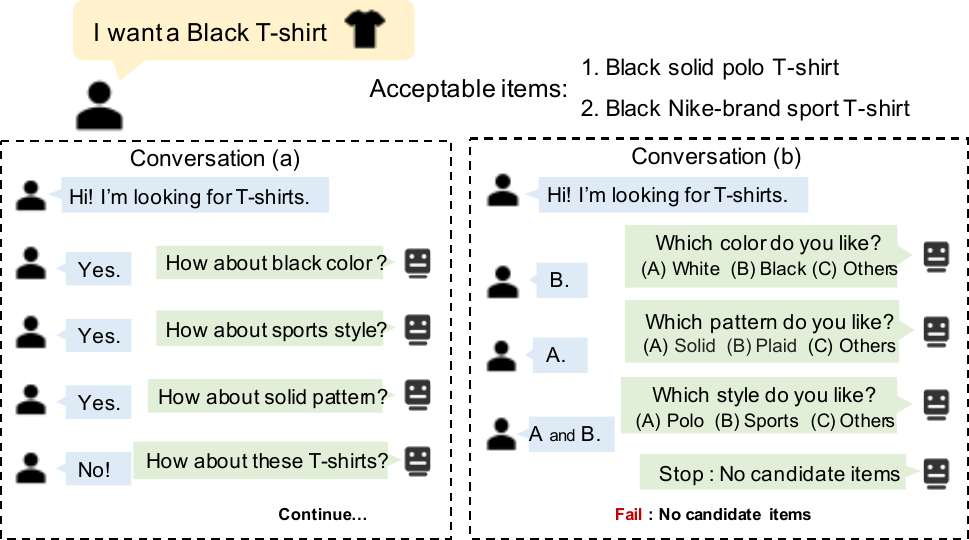}
    \vspace{-0.6cm}
    \caption{Examples of MIMCR scenario.}
    \label{example}
    \vspace{-0.6cm}
\end{figure}

To this end, we extend the MCR to a more realistic scenario, namely Multi-Interest Multi-round Conversational Recommendation (MIMCR), in which users may have multiple interests in attribute instance combinations and accept multiple items with partially overlapped combinations of attribute instances. 
As shown in Figure~\ref{example}, the user wants a black T-shirt. For the attribute types such as "style" or "brand", he can accept one or more instances. He shows interest in the combinations of "Nike-brand" and "sports", as well as "solid" and "polo" respectively. The user could accept a "black solid polo" T-shirt or a "black Nike-brand sports" T-shirt.
The task will be completed as CRS successfully recommends one of them. 

        
        
Existing works may encounter three significant limitations under the MIMCR scenario. First, current CRS frameworks often employ binary questions \cite{EAR}, which is concise but unable to elicit user interests effectively. 
As shown in the conversation (a) in Figure~\ref{example}, although the user accepts all of the attribute instances asked by CRS, the combination of them does not point to any target items the user prefers. 
Moreover, since the CRS agent has asked attribute instance "sports", it will hardly ask "polo" (the user favors). This is the result of the mutual exclusion of attribute instances with the same attribute type in the current CRS system design. 
On the other hand, enumerating all choices (associated with each attribute instances) \cite{EAR,KBQG,zou2020towards} are not practical since there may be too many attribute instances to be shown and answered by the user.
Second, as shown in the conversation (b) in Figure~\ref{example}, CRS can efficiently obtain user preferences by using multiple choice questions. However, the existing methods utilize the intersection set strategy to select items that are associated with all accepted attribute instances, which could easily lead to the over-filter of user preferred candidate items as the conversation progresses. 
Finally, the existing methods simply model user's intentions in a uniform manner, while neglecting the diversity of user interests, which will often fail to identify the user's multiple interests through the combinations of attribute instances.


To effectively address the aforementioned challenges, we propose a novel framework named Multiple Choice questions based Multi-Interest Policy Learning (MCMIPL) for MIMCR. 
In order to obtain user preferences more efficiently, our method generates attribute type-based multiple choice questions. As the conversation (b) in Figure~\ref{example}, the user can flexibly select the attribute instances he likes or the option "Others" if he likes none. 
To avoid over-filtering items, we propose a union set strategy to select candidate items. In particular, we select the items satisfying at least one of the accepted attribute instances as the candidate items. 
Moreover,  we develop a Multi-Interest Policy Learning (MIPL) module to decide the next action, either asking or recommending items. In details, we construct a current graph based on the conversation state, and a global graph based on the historical user-item interactions and the global item-attribute instance correlations. 
Based on the representation learned by graph neural network (GNN), we iteratively capture multiple interests of the user. Finally, the next action will be decided based on the policy learning with the multi-interest representations.


The contributions of this work are summarized as follows:
\begin{itemize}[leftmargin=*]
    \item We extend existing CRS to a more realistic scenario setting named MIMCR, which comprehensively takes into account the incompleteness and diversity of user's interests.
    \item For the MIMCR scenario, we propose the MCMIPL framework with more appropriate strategies to generate questions and select candidate items. Furthermore, our method iteratively extracts the user's multiple interests based on the current state and historical global information, to decide the next action via policy learning. 
    \item We adapt four datasets for MIMCR, and extensive experimental results on these datasets show the superiority of our method.
\end{itemize}

\section{Related Works}
\subsection{Conversational Recommendation}
Compared to existing sequential or social recommendation systems \cite{pang2021heterogeneous,zhang2021graph,shen2021multi}, Conversational Recommendation System (CRS) is an effective solution for dynamic user preference modeling and explainable recommendation, originated from task-oriented dialogue systems \cite{lei2018sequicity}.
Through the conversations with users, CRS collects the user’s preference and then generates recommendations directly.
In recent years, various approaches  \cite{lewis2017deal,lee2019melu,chen2019KRRD,zhou2020KGSF,xie2021comparison,priyogi2019preference} based on deep learning and reinforcement learning (RL) have been proposed for CRS.
Multi-Armed Bandits based methods \cite{Abs_Greedy,li2021seamlessly,zhang2018towards} and meta-learning based methods \cite{lee2019melu,zou2020neural} solve the user cold-start problem and balance the exploration and exploitation trade-offs for CRS. 
Besides, some methods \cite{zhao2013interactive,sepliarskaia2018preference,zou2020pseudo} focus on asking questions about items to obtain the users' preference.
In addition, the approaches focusing on the dialogue ability \cite{li2018TDCR,chen2019KRRD,zhou2020KGSF}, are more likely to understand user's preferences and intentions with the input of raw natural language, and automatically generate fluent responses.

The most realistic conversational recommendation setting proposed so far is multi-round conversational recommendation (MCR) \cite{EAR,SCPR,Unicorn,xu2021adapting}. In MCR task, the system focuses on whether to ask attributes or make recommendations based on policy learning at each turn to hit the target item for fewer interaction turns to improve the user experience. In this work, we focus on the MCR problem.

\subsection{Multi-round Conversational Recommendation}

For multi-round conversational recommendation, a conversation strategy is essential in the interaction process.
The key of the conversation strategy is to dynamically decide when to ask questions, and when to make recommendations.
At current stage of research, several reinforcement learning (RL) based frameworks have been adopted into MCR to model the complex conversational interaction environment.
For instance, EAR \cite{EAR} utilizes latent vectors based on available information to capture the current state of MCR, and learns the proper timing to ask questions about attributes or to recommend.
Furthermore, SCPR \cite{SCPR} models the MCR task as an interactive path reasoning problem on the knowledge graph (KG).
It chooses attributes and items strictly following the paths, and reasons on KG to find the candidate attributes or items via user's feedback.
KBQG \cite{KBQG} generates the clarifying questions to collect the user’s preference of attribute types based on knowledge graph.
UNICORN \cite{Unicorn} proposes a unified reinforcement learning framework based on dynamic weighted graph for MCR, which unifies three decision-making processes.
Moreover, some sophisticated conversational strategies try to lead dialogues \cite{wu2019proactive}, which can introduce diverse topics and tasks in MCR \cite{lewis2017deal,wang2019Good,liu2020MT,wu2019chat,zhou2020Topic,wang2021finetuning,wong2021improving}.

However, these works all ignore a more realistic scenario in which users may accept multiple items with partially overlapped attributes. Therefore, we propose a new scenario named MIMCR to fill this gap. Furthermore, we develop a novel framework namely MCMIPL to tackle the existing challenges.
\section{DEFINITION AND PRELIMINARY}
Although the multi-round conversational recommendation (MCR) scenario \cite{EAR,SCPR,Unicorn} is the most realistic CRS setting proposed so far, the assumption proposed by MCR \cite{EAR}, that the user preserves clear preferences towards all the attributes and items, still deviates from real scenario. 
In this work, we assume the user's preference for items is incomplete when resorting to CRS. Specifically, the user has clear single preferences for some attribute types, while for other attribute types, his preference might be various or vague. With the guidance of CRS, he may accept multiple attribute instances with the same type, which results in that the user may show interests in over items under different attribute instance combinations. Therefore, we propose a new scenario named \textbf{M}ulti-\textbf{I}nterest \textbf{M}ulti-round \textbf{C}onversational \textbf{R}ecommendation (MIMCR).

In this scenario, we define the sets of users and items as $\mathcal{U}$ and $\mathcal{V}$, respectively. And we also separately define the sets of attribute types and instances as  $\mathcal{C}$ and $\mathcal{P}$. Each $v \in \mathcal{V}$ is associated with a set of attribute instances $\mathcal{P}_v$. 
Each  $p \in \mathcal{P}$ has its corresponding attribute type $c_p \in \mathcal{C}$. In each episode, there is a set $\mathcal{V}_u$ of items that are acceptable to the user $u \in \mathcal{U}$. The set is represented as follows:
\begin{equation}
\setlength{\abovedisplayskip}{3pt}
\setlength{\belowdisplayskip}{3pt}
\begin{aligned}
    \mathcal{V}_u=\{v_1,v_2,...v_{N_v}\}
\end{aligned}\label{N_t}
\end{equation}
where ${N_v}$ is the number of acceptable items, $\mathcal{P}_1 \cap \mathcal{P}_2 \cap\dots\cap \mathcal{P}_{N_v}=\mathcal{P}_{same}\neq \emptyset$ and $\mathcal{P}_{i} \neq \mathcal{P}_j$.
A conversation session is initialized by user $u$ specifying an attribute instance $p_0\in \mathcal{P}_{same}$ he clearly prefers. Then, the agent selects to ask questions about attribute instances or to recommend items based on policy learning. The CRS will update the conversational state based on the user feedback. The process will repeat until at least one acceptable item is successfully recommended to the user or the system reaches the maximum number of turn $T$.


\begin{figure*}[t]
    \vspace{-0.4cm}
    \centering
    \includegraphics[width=0.8\textwidth]{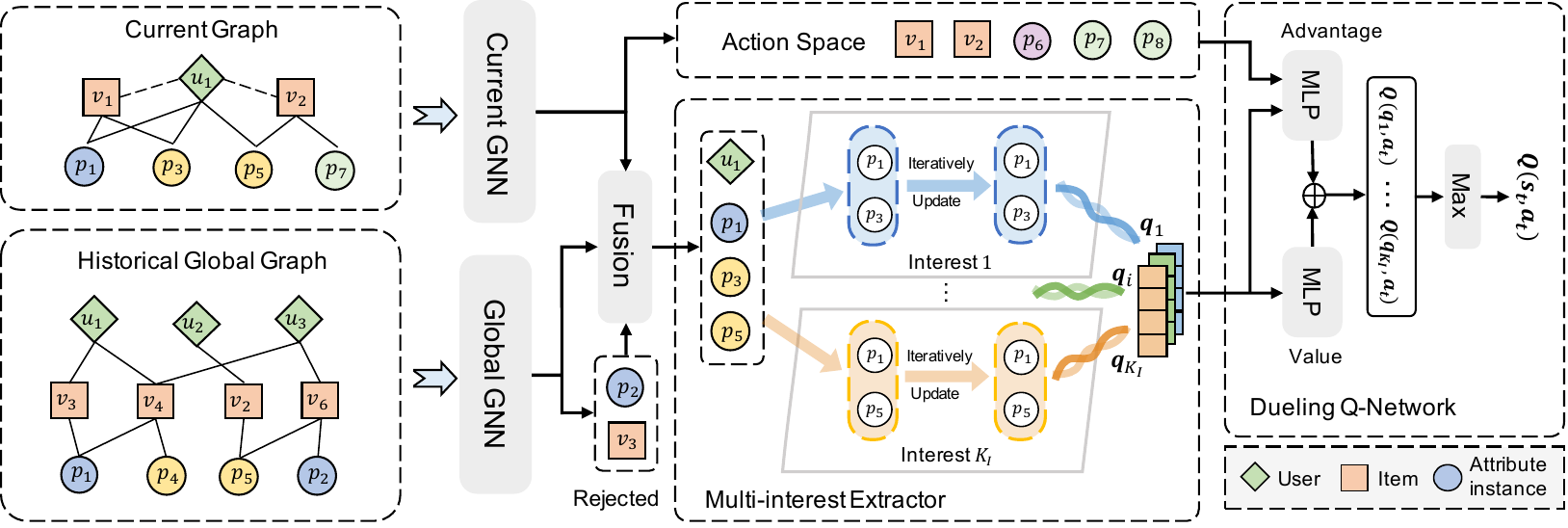}
    \vspace{-0.3cm}
    \caption{The overview of Multi-Interest Policy Learning (MIPL).}
    \label{fig:MIPL}
    \vspace{-0.4cm}
\end{figure*}


\section{Framework}
We propose Multiple Choice questions based Multi-Interest Policy Learning (MCMIPL), a novel framework for MIMCR. The goal of our framework is to learn the policy network $\pi(s_t|a_t)$ to maximize the expected cumulative rewards as: $\pi^*={\rm argmax}_{\pi\in\Pi}\mathbb{E}\left[\sum_{t=0}^{T}r_t\right]$,
where $s_t$ denotes the current state, $a_t$ denotes the  action taken by the agent and the $r_t$  is intermediate reward.
On the whole, the process of our framework in one turn can be decomposed into three steps: user modeling, consultation and transition.

\subsection{User Modeling}
We firstly encode the state $s_t$, which contains all the conversational information of the prior $t-1$ turns. The current state includes six components: $s_t=\{u,\mathcal{P}^{(t)}_u,\mathcal{P}^{(t)}_{rej},\mathcal{V}^{(t)}_{rej},\mathcal{P}^{(t)}_{cand},\mathcal{V}^{(t)}_{cand}\}$.
Previous methods \cite{EAR,SCPR,Unicorn} for MCR only extract the user's interest from the current state, ignoring the complements of historical interactions to the current user's preference. To this end, we construct a current graph and a global graph to jointly learn user representations. Moreover, we develop an iterative multi-interest extractor to obtain multiple interests of the user, which will be discussed in \autoref{multi-interest}.

\subsection{Consultation}
Once the system finishes the user modeling step, it will move to the consultation step, with the purpose  to decide whether to ask attribute instances or to recommend items. 
To make the next action more profitable and recommend successfully with the fewer turns, we employ a reinforcement learning (RL) method based on the extracted multiple interests of the user to learn the policy. 
The action space includes all candidate items and candidate attribute instances. However, in the real world, the number of items and attribute instances is very large, which severely limits the efficiency of CRS. To improve the efficiency, we sample $K_v$ items and $K_p$ attribute instances as action space $\mathcal{A}_{t}$. We develop a novel dueling Q-network \cite{wang2016dueling} to calculate the Q-value of each action in $\mathcal{A}_{t}$. If CRS decides to ask a question, 
our method will select $K_a$ attribute instances in $\mathcal{A}_{t}$ with the same attribute type to generate \emph{attribute type-based multiple choice questions}. 
The user can choose zero (the option "Others" as shown in conversation (b) of \autoref{example}), one, or more attribute instances with the given attribute type. If CRS decides to recommend items, the system will select $K$ items in $\mathcal{A}_{t}$ to recommend. We will discuss the details of sampling 
strategies and policy learning in \autoref{policy-learning}.

\subsection{Transition} \label{transition}
When the user responds to the action of agent, the transition step will be triggered. This step will transition the current state to the next state $s_{t+1}$. If the user responds to the question, attribute instance sets that the user accepts and rejects in this turn can be defined as  $\mathcal{P}_{cur\_acc}^{(t)}$ and $\mathcal{P}_{cur\_rej}^{(t)}$  respectively. Some components are updated by $\mathcal{P}_{cand}^{(t+1)}=\mathcal{P}_{cand}^{(t)}-\mathcal{P}_{cur\_rej}^{(t)}-\mathcal{P}_{cur\_acc}^{(t)}$, $\mathcal{P}_{rej}^{(t+1)}=\mathcal{P}_{rej}^{(t)}\cup \mathcal{P}_{cur\_rej}^{(t)}$ and $\mathcal{P}_{u}^{(t+1)}=\mathcal{P}_{u}^{(t)}\cup \mathcal{P}_{cur\_acc}^{(t)}$. When the user is recommended items, if the set $\mathcal{V}_{rec}^{(t)} $ of recommended items are all rejected, the next state can be updated by $\mathcal{V}_{rej}^{(t+1)}=\mathcal{V}_{rej}^{(t)}\cup \mathcal{V}_{rec}^{(t)}$. Otherwise, this conversation session ends. 
Finally, we need to update the candidate item set $\mathcal{V}_{cand}^{(t+1)}$ based on the user's feedback. Previous works \cite{SCPR,Unicorn} update candidate items based the intersection set strategy, that is, only the items satisfying all the accepted attribute instances in $\mathcal{P}_{u}^{(t+1)}$ remain, which obviously deviates from the scenario. In fact, the user might not prefer the combination of all attribute instances, but rather part of them. To this end, we propose the \emph{attribute instance-based union set strategy} to update $\mathcal{V}_{cand}^{(t+1)}$ as follows:
\begin{equation}
\setlength{\abovedisplayskip}{3pt}
\setlength{\belowdisplayskip}{3pt}
\begin{aligned}
    \mathcal{V}_{cand}^{(t+1)}=\{v| v\in\mathcal{V}_{p_0}-\mathcal{V}_{rej}^{(t+1)}&\ \ {\rm and}\ \   \mathcal{P}_{v} \cap  \mathcal{P}_{u}^{(t+1)}\neq \emptyset \\
    &\ \ {\rm and}\ \   \mathcal{P}_{v} \cap  \mathcal{P}_{rej}^{(t+1)}= \emptyset\}
\end{aligned}\label{N_t}
\end{equation}
where $\mathcal{V}_{p_0}$ is the item set in which all items are associated to attribute instance $p_0$ which initializes the conversation session. In this way, we can get the next state, which will be updated as $ s_{t+1}=\{u,\mathcal{P}^{(t+1)}_u,\mathcal{P}^{(t+1)}_{rej},\mathcal{V}^{(t+1)}_{rej},\mathcal{P}^{(t+1)}_{cand},\mathcal{V}^{(t+1)}_{cand}\}$.

\subsection{Reward} \label{reward}
In this work, five kinds of rewards are defined following \cite{SCPR,Unicorn}, namely, (1) $r_{rec\_suc}$, a strongly positive reward when the recommendation succeeds,
(2) $r_{rec\_fail}$, a strongly negative reward when the recommendation fails, 
(3) $r_{ask\_suc}$, a slightly positive reward when the user accepts an asked attribute instance,
(4) $r_{ask\_fail}$, a negative reward when the user rejects an asked attribute instance,
(5) $r_{quit}$, a strongly negative reward if the session reaches the maximum number of turns. 
In addition, since our method asks multiple choice questions, we design the reward  from the user's feedback on a question in the form of sum as $r_t=\sum_{ \mathcal{P}_{cur\_acc}^{(t)}} r_{ask\_suc}+\sum_{ \mathcal{P}_{cur\_rej}^{(t)}} r_{ask\_rej}$.



\section{Multi-interest Policy Learning}
In this section, we detail the design of Multi-Interest Policy Learning (MIPL) module. As shown in \autoref{fig:MIPL}, to obtain more comprehensive user representations, we establish a current graph to capture user current preferences, and a global graph to capture long-term preferences. Based on the learned node representations of the two graphs, we propose an iterative multi-interest extractor to model user's preferences for different combinations of attribute instances. Moreover, we design a new dueling Q-network \cite{wang2016dueling} to decide the next action based on the extracted multiple interests.
\subsection{Multi-interest Encoder} \label{multi-interest}
\subsubsection{GNN-based Representation Fusion}
The existing methods \cite{EAR,SCPR,Unicorn} capture user preferences based on the current conversation state, which might cause user preferences to be incomplete due to the limited number of turns. In addition, only the current conversation information is not enough to capture the correlation of attribute instances. Therefore, we construct a current graph based on the conversation state, and a historical global graph based on the historical user-item interactions and the global item-attribute instance correlations. We employ GNNs to learn the node representations of two graphs separately and utilize gating mechanism for fusion.

\textbf{Current Graph Representation}.
Following \cite{Unicorn}, we construct a weighted graph based on the $t$-th turn state of a episode as $\mathcal{G}_u^{(t)}=(\mathcal{N}^{(t)},\mathcal{E}^{(t)})$, where $\mathcal{N}^{(t)}=\{u\} \cup \mathcal{P}^{(t)}_u\ \cup \mathcal{P}^{(t)}_{cand} \cup \mathcal{V}^{(t)}_{cand}$. For the edge weight $\mathcal{E}^{(t)}_{i,j}$, we consider three cases: (1) The weight of edge between the user and each accepted attribute instance is $1$; (2) The weight of edge between each attribute instance and the associated item   is $1$; (3) The weight of edge between the user and each item is $w_v^{(t)}$, which indicates the coarse matching score of the item $v$ to the current state: $w_v^{(t)}=\sigma(\mathbf{e}_u^T \mathbf{e}_v+\sum_{p\in \mathcal{P}^{(t)}_u} \mathbf{e}_v^T \mathbf{e}_p-\sum_{p\in \mathcal{P}^{(t)}_{rej}} \mathbf{e}_v^T \mathbf{e}_p)$,
where $\sigma(\cdot)$ is the sigmoid function, $\mathbf{e}_u$, $\mathbf{e}_v$ and $\mathbf{e}_p \in \mathbb{R}^d$ are the initial embedding of user, item and attribute instance.

We employ a $L_c$-layer GCN \cite{kipf2016semi} to capture the connectivity between nodes of $\mathcal{G}_u^{(t)}$ and obtain higher-quality node representations in the current state. We define the initial embedding $\mathbf{e}_n$ of node $n$ as $\mathbf{e}_n^{(0)}$ , and $\mathbf{e}_n^{(l)}$ as the output node embedding of $l$-th layer. The calculation method of $l+1$-th layer is as follows:
\begin{equation}
\setlength{\abovedisplayskip}{3pt}
\setlength{\belowdisplayskip}{3pt}
\begin{aligned}
     \mathbf{e}^{(l+1)}_{n}={\rm ReLU}(\sum_{j\in \mathcal{N}^{(t)}_n}{\frac{\mathbf{W}_c^{(l+1)}\mathbf{e}^{(l)}_j}{\sqrt{\sum_i \mathcal{E}^{(t)}_{n,i} \sum_i \mathcal{E}^{(t)}_{j,i}}}}+\mathbf{e}^{(l)}_n)
\end{aligned}\label{GCN current}
\end{equation}
where $\mathcal{N}^{(t)}_n$ denotes the set of neighbor nodes of node $n$ in the turn~$t$, $\mathbf{W}_c^{(l+1)} \in \mathbb{R}^{d\times d}$ is trainable parameters. We define the output of the last layer $\mathbf{e}^{(L_c)}_{n}$ as the final embedding $\mathbf{e}^{c}_{n}$ of the node.

\textbf{Global Graph Representation}.
We use the historical interactions between users and items as well as the correlation between items and attribute instances to establish a heterogeneous global graph $\mathcal{G}_g=(\mathcal{N},\mathcal{E})$, where
$\mathcal{N}=\mathcal{U} \cup \mathcal{V} \cup \mathcal{P}$ and $\mathcal{E}=\mathcal{E}_{u,v} \cup \mathcal{E}_{p,v}$.
The edge $(u,v,r_{u \sim v}) \in \mathcal{E}_{u,v}$ denotes the user $u$ has interacted the item $v$. And the edge $(p,v,r_{p \sim v}) \in \mathcal{E}_{p,v}$ denotes that the item $v$ is associated with the attribute instance $p$.

We employ a $L_g$-layer  Global Graph Neural Network (GGNN) \cite{chen2019reinforcement,chen2020iterative,schlichtkrull2018modeling} to extract long-term interests of users, and global correlations of items and attribute instances.  The initial input embeddings of the first layer are $\mathbf{s}^{(0)}_u=\mathbf{e}_u$, $\mathbf{s}^{(0)}_v=\mathbf{e}_v$ and $\mathbf{s}^{(0)}_p=\mathbf{e}_p$. Let $\mathbf{s}^{(l)}_u$, $\mathbf{s}^{(l)}_v$ and $\mathbf{s}^{(l)}_p$ denote the output  representations of nodes after the propagation of $l$-th layer. For the $l+1$-th layer of GGNN, we model different edge types separately. For the edge in $\mathcal{E}_{u,v}$, we adopt the calculation method as follow:
\begin{equation}
\setlength{\abovedisplayskip}{3pt}
\setlength{\belowdisplayskip}{3pt}
\begin{aligned}
    \mathbf{s}_{u \sim v}^{(l+1)}(n)=\mathbf{b}_g^{(l+1)}+\sum_{i\in \mathcal{N}_{r_{u \sim v}}(n)}{\frac{\mathbf{W}^{(l+1)}_g\mathbf{s}^{(l)}_i}{\sqrt{\left| \mathcal{N}_{r_{u \sim v}}(i) \right| \left| \mathcal{N}_{r_{u \sim v}}(n)\right|}}}
\end{aligned}\label{HGGNN}
\end{equation}
where $\mathcal{N}_{r_{u \sim v}}(n)$ denotes the neighbor nodes of node $n$ with the edge type $r_{u \sim v}$, $\mathbf{W}_g^{(l+1)}$ and $\mathbf{b}_g^{(l+1)}$ are trainable parameters. For the edge in $\mathcal{E}_{p,v}$, we adopt the same method as Equation \ref{HGGNN} to get $\mathbf{s}_{p \sim v}^{(l+1)}(n)$.


For the user $u$ and attribute instance $p$, we utilize ReLU function to activate semantic messages to obtain output node embeddings: $\mathbf{s}^{(l+1)}_{u}={\rm ReLU}(\mathbf{s}_{u \sim v}^{(l+1)}(u))$, $\mathbf{s}^{(l+1)}_{p}={\rm ReLU}(\mathbf{s}_{p \sim v}^{(l+1)}(p))$.
Since item $v$ is connected by both two kinds of edges, we accumulate different messages propagated by different types of edges and update the representation: $\mathbf{s}^{(l+1)}_{v}={\rm ReLU} (\text{mean}(\mathbf{s}_{u \sim v}^{(l+1)}(v),\mathbf{s}_{p \sim v}^{(l+1)}(v)))$.
We define the output of the last layer $\mathbf{s}^{(L_g)}_{n}$ as the final node embedding $\mathbf{s}^{g}_{n}$.

We apply the gating mechanism to fuse the embeddings of nodes which belong to both graphs $\mathcal{G}_u^{(t)}$ and $\mathcal{G}_g$ as follows:
\begin{equation}
\setlength{\abovedisplayskip}{3pt}
\setlength{\belowdisplayskip}{3pt}
\begin{aligned}
    g=\sigmoid(\mathbf{W}_{gated}\left[\mathbf{e}^{c}_{n} \parallel \mathbf{s}^{g}_{n}\right]),\ 
    \mathbf{v}_{n}=g \cdot \mathbf{s}^{g}_{n} + \left(1 - g\right) \cdot \mathbf{e}^{c}_{n},
\end{aligned}
\end{equation}
where $\parallel$ is the concatenate operation,  $\mathbf{W}_{gated}\in \mathbb{R}^{d\times d}$ is trainable parameter and  $\sigmoid(\cdot)$ is the sigmoid function.
    

\subsubsection{Iterative Multi-interest Extractor}
In CRS scenario, since the user's interest is diversity, we use multi-attention mechanism to model the user $u$ and attribute instances accepted by $u$. The multi-interest embeddings of user can be obtained through the combination of attribute instances with different weights. Inspired by \cite{capsule,cen2020controllable,wang2020disentangled}, we adopt the iterative update rule to adjust the weights of attribute instances with $M$ iterations more precisely.

Previous works rarely consider items or attribute instances rejected by users, which can complement the current preferences of the user effectively. Therefore, we first fuse the global embeddings of the rejected items and attribute instances with the user's embedding:
\begin{equation}
\setlength{\abovedisplayskip}{3pt}
\setlength{\belowdisplayskip}{3pt}
\begin{aligned}
    \mathbf{\hat{v}}_{u}=\mathbf{v}_{u}+\mathbf{W}_u(\frac{1}{\left| \mathcal{N}_{rej}\right|} \sum_{n\in\mathcal{N}_{rej}}\mathbf{s}^{g}_{n})
\end{aligned}\label{user_em}
\end{equation}
where $\mathcal{N}_{rej}=\mathcal{V}_{rej} \cup \mathcal{P}_{rej}$, and $\mathbf{W}_u\in \mathbb{R}^{d\times d}$ is trainable parameters. Then, we define $K_I$ attention networks 
for $K_I$ interests. Based on accepted attribute instance embeddings $[\mathbf{v}_{1},\mathbf{v}_{2},...,\mathbf{v}_{N}]$ and user embedding $\mathbf{\hat{v}}_{u}$, the initial iteration calculation method of each attention network to obtain the interest embedding $\mathbf{q}_k^{(1)}$ is as follows:
\begin{equation}
\setlength{\abovedisplayskip}{4pt}
\setlength{\belowdisplayskip}{4pt}
\begin{aligned}
    \mathbf{q}_k^{(1)}=\sum_{n=1}^N \alpha_{k,n}^{(1)} \mathbf{v}_{n},\ k\in\{1,...K_I\}
\end{aligned}\label{multi-interst}
\end{equation}
\begin{equation}
\setlength{\abovedisplayskip}{4pt}
\setlength{\belowdisplayskip}{4pt}
\begin{aligned}
    \alpha_{k,n}^{(1)}=\frac{{\rm exp}(\mathbf{h}_k^T\sigmoid(\mathbf{W}_k(\mathbf{\hat{v}}_{u}\parallel \mathbf{v}_{n})))}{\sum_{n'=1}^N {\rm exp}(\mathbf{h}_k^T\sigmoid(\mathbf{W}_k(\mathbf{\hat{v}}_{u}\parallel \mathbf{v}_{n'})))}
\end{aligned}\label{multi-interst_alpha}
\end{equation}
where $ \mathbf{h}_k$ and $ \mathbf{W}_k$ are trainable metrics. The $m$-th iteration precisely adjusts the weights $\alpha_{k,n}^{(m)}$ based on the $m-1$-th iteration results:
\begin{equation}
\setlength{\abovedisplayskip}{3pt}
\setlength{\belowdisplayskip}{3pt}
\begin{aligned}
    \mathbf{q}_k^{(m)}=\sum_{n=1}^N \alpha_{k,n}^{(m)} \mathbf{v}_{n}
\end{aligned}\label{multi-interst}
\end{equation}
\begin{equation}
\setlength{\abovedisplayskip}{3pt}
\setlength{\belowdisplayskip}{3pt}
\begin{aligned}
    \alpha_{k,n}^{(m)}=\frac{{\rm exp}(\mathbf{h}_k^T\sigmoid(\mathbf{W}_k(\mathbf{q}_k^{(m-1)}\parallel \mathbf{v}_{n}))+\alpha_{k,n}^{(m-1)})}{\sum_{n'=1}^N {\rm exp}(\mathbf{h}_k^T\sigmoid(\mathbf{W}_k(\mathbf{q}_k^{(m-1)}\parallel \mathbf{v}_{n'}))+\alpha_{k,n'}^{(m-1)})}
\end{aligned}\label{multi-interst_alpha}
\end{equation}
where $ \mathbf{h}_k$ and $ \mathbf{W}_k$ are parameters shared with the previous iterations. We define the output $\{\mathbf{q}_1^{(M)},\mathbf{q}_2^{(M)},...,\mathbf{q}_{K_I}^{(M)}\}$ of $M$-th iteration as the final multi-interest embeddings $\{\mathbf{q}_1,\mathbf{q}_2,...\mathbf{q}_{K_I}\}$.

\subsection{Action Decision Policy Learning} \label{policy-learning}
A large action search space will bring a great negative impact on the efficiency of the system. Following \cite{Unicorn}, we select $K_v$ items and $K_p$ attribute instances as candidate action space $\mathcal{A}_{t}$. For candidate items to be recommended, we consider how well they match the current state. We select top-$K_v$ items into the action space based on $w_v^{(t)}$. For attribute instances, we also select top-$K_p$ attribute instances based on $w_p^{(t)}$
as: $w_p^{(t)}=\sigma(\mathbf{e}_u^T \mathbf{e}_p+\sum_{p'\in \mathcal{P}^{(t)}_u} \mathbf{e}_p^T \mathbf{e}_{p'}-\sum_{p\in \mathcal{P'}^{(t)}_{rej}} \mathbf{e}_p^T \mathbf{e}_{p'})$

Inspired by \cite{Unicorn}, we design an improved dueling Q-network \cite{wang2016dueling} to determine the next action. Following the standard assumption that delayed rewards are discounted by a factor of $\gamma$ per timestep, we define the Q-value $Q(s_t,a_t)$ as the expected reward based on the state $s_t$ and the action $a_t$. Based on the obtained $K_I$ interest representations according to the current state $s_t$, we calculate each score between action $a_t$ and each interest, and take the maximum value as Q-value:
\begin{equation}
\setlength{\abovedisplayskip}{3pt}
\setlength{\belowdisplayskip}{3pt}
\begin{aligned}
    Q(s_t,a_t)=\max_k( f_{\theta_V}(\mathbf{q}_k)+f_{\theta_A}(\mathbf{q}_k,a_t)),\ k\in\{1,...K_I\}
\end{aligned}\label{Q-learning}
\end{equation}
where $f_{\theta_V}(\cdot)$ and $f_{\theta_A}(\cdot)$ are separate multi-layer perceptions (MLP). The optimal Q-function $Q^*(s_t, a_t)$ achieves the maximum expected reward by the optimal policy $\pi^*$, following the Bellman~\cite{bellman1957role} equation:
\begin{equation}
\setlength{\abovedisplayskip}{3pt}
\setlength{\belowdisplayskip}{3pt}
\begin{aligned}
    Q^*(s_t,a_t)=\mathbb{E}_{s_{t+1}}\left[ r_t+\gamma \max_{a_{t+1}\in \mathcal{A}_{t+1}} Q^*(s_{t+1},a_{t+1}| s_t,a_t)\right]
\end{aligned}\label{score}
\end{equation}

The CRS firstly selects the action with the max Q-value. If the selected action points to an item, the system will recommend top-$K$ items with the highest Q-value to the user. If the selected action points to an attribute instance $p$, the system will generate \emph{attribute type-based multiple choice questions} to ask user. To be specific, the system will decide a attribute type $c$, and select top-$K_a$ attribute instances whose corresponding attribute type is $c$ with the highest Q-value. Then the user can choose which of the attribute instances he likes or dislikes. We propose two strategies to decide the attribute type $c$: (1) Top-based strategy. We select the attribute type corresponding to the attribute instance with the highest Q-value. (2) Sum-based strategy. For each attribute type, we sum the Q-values of its corresponding attribute instances to obtain the attribute type level score, and select the attribute type with the highest score. During the experiment, we mainly use Top-based strategy, and the other strategy will be compared in the ablation study.

\subsection{Model Training}
For each turn, the agent will receive the reward $r_t$ based on the user's feedback. According to user feedback, we can update the state $s_{t+1}$ and action space $\mathcal{A}_{t+1}$. We define a replay buffer $\mathcal{D}$ following \cite{Unicorn}, which stores the experience $(s_t,a_t,r_t,s_{t+1},\mathcal{A}_{t+1})$. To train our model, we sample mini-batch experiences from the replay buffer~$\mathcal{D}$ and define a loss function as follows:
\begin{equation}
\setlength{\abovedisplayskip}{3pt}
\setlength{\belowdisplayskip}{3pt}
\begin{aligned}
    \mathcal{L}=\mathbb{E}_{(s_a,a_t,r_t,s_{t+1},\mathcal{A}_{t+1})\sim\mathcal{D}}\left[(y_t-Q(s_t,a_t;\theta_Q,\theta_M))^2\right]
\end{aligned}\label{score}
\end{equation}
where $\theta_M$ is the set of parameters to capture multi-interest embeddings, $\theta_Q=\{\theta_V,\theta_A\}$, and $y_t$ is the target value, which is based on the optimal Q-function as follows:
\begin{equation}
\setlength{\abovedisplayskip}{3pt}
\setlength{\belowdisplayskip}{3pt}
\begin{aligned}
    y_t=r_t+\gamma \max_{a_{t+1}\in \mathcal{A}_{t+1}} Q(s_{t+1},a_{t+1};\theta_Q,\theta_M)
\end{aligned}\label{score}
\end{equation}

Due to the overestimation bias in original DQN, we employ the double DQN \cite{van2016deep} to copy a target network $Q'$ as a periodic from the online network to train the model following \cite{zhou2020interactive,Unicorn}.
\section{Experiments}
To fully demonstrate the superiority of our method,
we conduct experiments to verify the following four research questions (RQ):
\begin{itemize}[leftmargin=*]
    \item \textbf{(RQ1)}: Compared with the state-of-the-art methods, does our framework achieve better performance?
    
    \item \textbf{(RQ2)}: What are the impacts of key components on performance?
    
    \item \textbf{(RQ3)}: How do the settings of hyper-parameters (such as the number of interests $K_I$) affect our framework?
    
    \item \textbf{(RQ4)}: How can our framework effectively extract multiple interests in different attribute instance combinations?
    
\end{itemize}
\begin{table}[htbp]
    \vspace{-0.3cm}
    \caption{Statistics of datasets. }
    \vspace{-0.3cm}
    \label{tab:ws-darts}
    \centering
    \small
        \begin{tabular}{p{2.2cm} p{1.0cm}<{\raggedleft}p{1.0cm}<{\raggedleft}p{1.0cm}<{\raggedleft}p{1.2cm}<{\raggedleft}}  
    \toprule
    \textbf{Dataset}   & \textbf{Yelp} & \textbf{LastFM} &\textbf{Amazon-Book}&\textbf{MovieLens}\\
    \midrule
    $\#$Users&27,675&1,801&30,291&20,892 \\
    $\#$Items &70,311&7,432&17,739&16,482\\
    $\#$Interactions&1,368,609&76,693&478,099&454,011\\
    $\#$Attribute instances&590&8,438&988&1,498\\
    $\#$Attribute types&29&34&40&24\\
    \midrule
    $\#$Entities&98,576&17,671&49,018&38,872\\
    $\#$Relations&3&4&2&2\\
    $\#$Triplets&2,533,827&228,217&565,068&380,016\\
    \bottomrule
    \end{tabular}
    \vspace{-0.4cm}
\end{table}
\subsection{Datasets}\label{sec:standalone}
To evaluate the proposed method, we adapt two existing
MCR benchmark datasets, named Yelp and LastFM. To evaluate our method more comprehensively, we also process two additional datasets. The statistics of these datasets are presented in Table \ref{tab:ws-darts}.
\begin{itemize}[leftmargin=*]
    \item \textbf{Yelp} and \textbf{LastFM} \cite{EAR}: For the Yelp, Lei et al. build a 2-layer taxonomy. We define the 29 first-layer categories as attribute types, and 590 second-layer categories as attribute instances. For the LastFM, we adopt original attributes as attribute instances. We utilize clustering to select 34 categories as attribute types.
    \item \textbf{Amazon-Book} ~\cite{wang2019kgat}: We select entities and relations in knowledge graph (KG) as attribute instances and types, separately. To ensure data quality, we select entities associated with at least 10 items.
    \item \textbf{MovieLens}: MovieLens-20M\footnote{https://grouplens.org/datasets/movielens/} is a widely used recommendation benchmark dataset. We retain the user-item interactions with the rating $> 3$. Similarly, we select entities in KG as attribute instances and relations as attribute types.
\end{itemize}
For each conversation episode, we sample $N_v$ items with partially overlapped attribute instances as the acceptable items for the user.

\subsection{Experiments Setup}

\subsubsection{User Simulator}
Since MCR is a system based on interaction with users, we design a user simulator to train and evaluate it. Based on the scenario MIMCR, we adjust the user simulator adopted in \cite{EAR}. We simulate a conversation session for each observed user-item set interaction pair $(u,\mathcal{V}_u)$. We regard each item $v_i \in \mathcal{V}_u$ as the ground-truth target item. The session is initialized by the simulated user specifying an attribute instance $p_0 \in \mathcal{P}_{same}$. Given a conversation, the simulated user's feedback of each turn follows the rules: (1) when the system asks a question, he will accept the attribute instances which are associated with any item in $\mathcal{V}_u$ and reject others; (2) when the system recommends a list of items, he will accept it if the list contains at least one item in $\mathcal{V}_u$; (3) We consider that user's patience will run out when the maximum number of turn $T$ is reached. The simulated user will exit the system until he accepts the recommended item list or his patience runs out.
\begin{table*}[t]
    \centering
    \caption{Performance comparison of different models on the four datasets. hDCG stands for hDCG@($15,10$).}
    \vspace{-0.3cm}
    \label{results}
    \begin{tabular}{p{2.0cm}<{\centering}p{0.8cm}<{\centering}p{0.8cm}<{\centering}p{0.8cm}<{\centering}p{0.01cm}p{0.8cm}<{\centering}p{0.8cm}<{\centering}p{0.8cm}<{\centering}p{0.01cm}p{0.8cm}<{\centering}p{0.8cm}<{\centering}p{0.8cm}<{\centering}p{0.01cm}p{0.8cm}<{\centering}p{0.8cm}<{\centering}p{0.8cm}<{\centering}}
    \toprule
    \multirow{2}{*}{\bfseries Models }&\multicolumn{3}{c}{\bfseries Yelp }&&\multicolumn{3}{c}{\bfseries LastFM }&&\multicolumn{3}{c}{\bfseries Amazon-Book }&&\multicolumn{3}{c}{\bfseries MovieLens }\\
    \cline{2-4}
    \cline{6-8}
    \cline{10-12}
    \cline{14-16}
    &SR@15&AT&hDCG&&SR@15&AT&hDCG&&SR@15&AT&hDCG&&SR@15&AT&hDCG\\
    \midrule
    Abs Greedy &0.195&14.08&0.069&&0.539&10.92&0.251&&0.214&13.50&0.092&&0.752& 4.94&0.481\\
    Max Entropy&0.375&12.57&0.139&&0.640&9.62&0.288&&0.343&12.21&0.125&&0.704& 6.93&0.448\\
    CRM& 0.223&13.83&0.073&&0.597&10.60&0.269&&0.309&12.47&0.117&&0.654&7.86&0.413\\
    EAR&0.263&13.79&0.098&&0.612&9.66&0.276&&0.354&12.07&0.132&&0.714&6.53&0.457\\
    SCPR&0.392&12.65&0.140&&0.659&9.36&0.307&&0.390&11.72&0.144&&0.799&4.39&0.529\\
    UNICORN& 0.404&12.39&0.146&&0.788&7.56&0.355&&0.416&11.68&0.155&&0.819&4.28&0.568\\
    \midrule
    $\text{SCPR}^{\dagger}$ &0.413&12.45&0.149&&0.751&8.52&0.339&&0.428&11.50&0.159&&0.812&4.03&0.547\\
    $\text{UNICORN}^{\dagger}$ &0.438&12.28&0.151&&0.843&7.25&0.363&&0.466&11.24&0.170&&0.836&3.82&0.576\\
    \midrule
    Our Model&{\bfseries 0.482}&{\bfseries 11.87}&{\bfseries 0.160}&&{\bfseries 0.874}&{\bfseries 6.35}&{\bfseries 0.396}&&{\bfseries 0.545}&{\bfseries 10.83}&{\bfseries 0.223}&&{\bfseries 0.882}&{\bfseries  3.61}&{\bfseries 0.599}\\
    
    \bottomrule
    \end{tabular}
    \vspace{-0.3cm}
\end{table*}

\subsubsection{Baselines}
To evaluate model performance, we compare our model with following six representative baselines:
\begin{itemize}[leftmargin=*]
    \item \textbf{Max Entropy} asks an attribute or recommends the top ranked items based on a certain probability \cite{EAR}.
    \item \textbf{Abs Greedy} \cite{Abs_Greedy} only recommends items in each turn and treats rejected items as negative examples to update the model.
    \item \textbf{CRM} \cite{CRM} is originally designed for single-round CRS, which utilizes reinforcement learning to select next action. Following \cite{EAR}, we adapt CRM to MCR scenario.
    \item \textbf{EAR} \cite{EAR} adopts a three stage solution called Estimation–Action–\\Reflection for MCR, and employs RL strategy to decide actions.
    \item \textbf{SCPR} \cite{SCPR} proposes a generic framework that models MCR as an interactive path reasoning problem on a graph, and employs the DQN \cite{DQN} framework to select actions.
    \item \textbf{UNICORN} \cite{Unicorn} proposes a unified policy learning framework, which develops a dynamic graph based RL to select action for each turn. It is the state-of-the-art (SOTA) method.
\end{itemize}
For a more comprehensive and fair performance comparison, we adapt SCPR and UNICORN as follows: (1) The system employs multiple choice questions to ask the user. When the system decides to ask the user, the agent will generate  \emph{attribute type-based multiple choice questions} as described in \autoref{policy-learning}. (2) The system selects candidate item set by the \emph{attribute instance-based union set strategy} described in \autoref{transition}. We name the two adapted methods $\text{SCPR}^{\dagger}$ and $\text{UNICORN}^{\dagger}$ respectively.

\subsubsection{Parameters Setting}
We randomly split each dataset for training, validation and test with the ratio of $7:1.5:1.5$. The embedding dimension is set as 64, while the batch size as $128$. We recommend top $K=10$ items or ask $k_a=2$ attribute instances in each turn. The maximum turn $T$ of conversation is set as $15$. 
We employ the Adam optimizer with the learning rate $1e-4$. Discount factor $\gamma$ is set to be $0.999$. Following \cite{Unicorn}, we adopt TransE \cite{TransE} via OpenKE \cite{OpenKE} to pretrain the node embeddings in the constructed KG with the training set. We construct the global graph based on the training set.
The numbers of current GNN layers $L_c$ and global GNN layers $L_g$ are set to be $2$ and $1$, respectively. We extract user's multiple interests with $M=2$ iterations.
For the action space, we select $K_p=10$ attribute instances and $K_v=10$ items.
To maintain a fair comparison, we adopt the same reward settings: $r_{rec\_suc}=1, r_{rec\_fail}=-0.1, r_{ask\_suc}=0.01, r_{ask\_fail}=0.1, r_{quit}=-0.3$. We set the maximum number $N_v$ of acceptable items as 2. Other settings are explored in the hyper-parameter analysis.
\begin{figure}[t]
    \centering
    \begin{subfigure}{0.48\linewidth}
        \includegraphics[width=\textwidth]{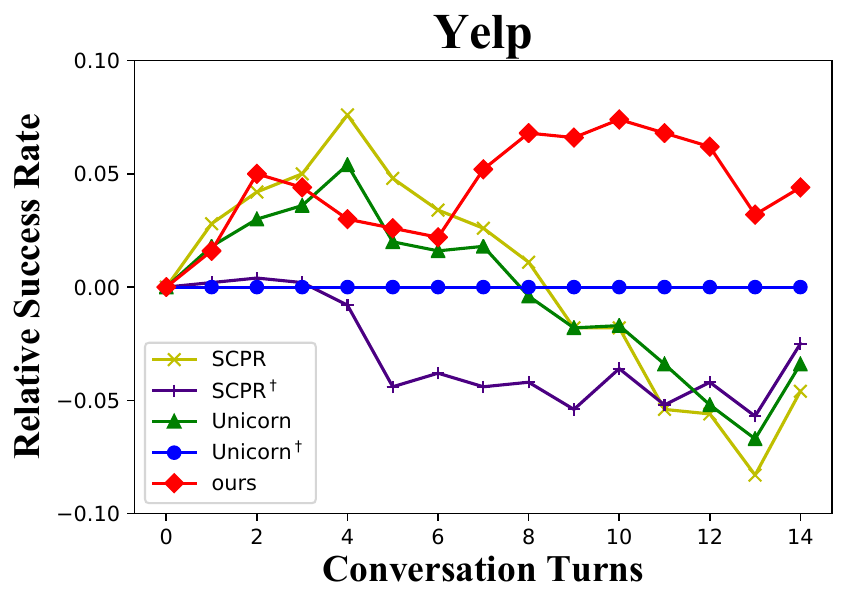}
    \end{subfigure}
    \begin{subfigure}{0.48\linewidth}
        \includegraphics[width=\textwidth]{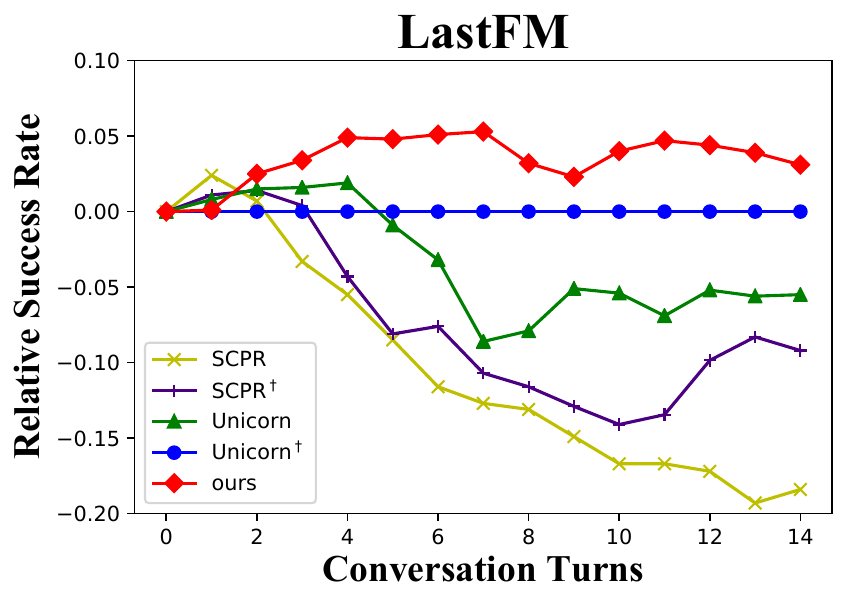}
    \end{subfigure}
    \begin{subfigure}{0.48\linewidth}
        \includegraphics[width=\textwidth]{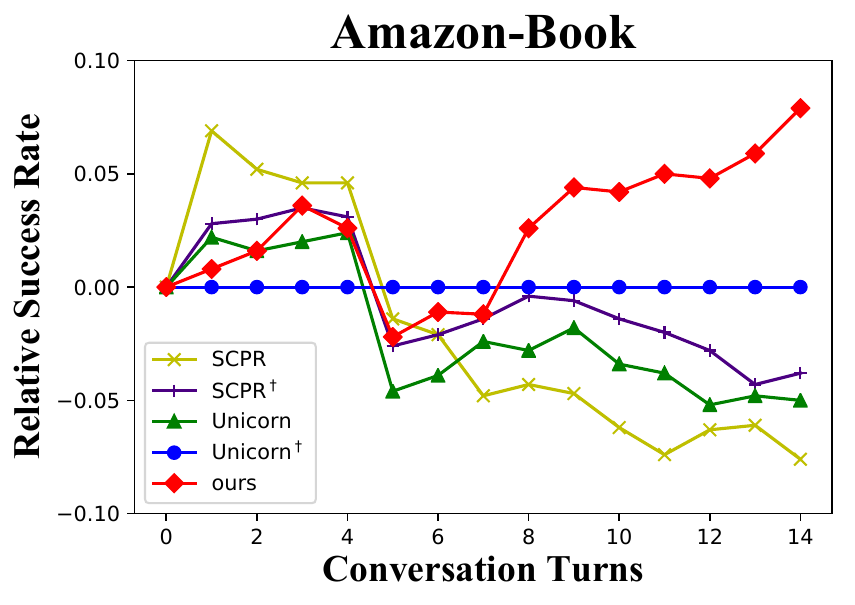}
    \end{subfigure}
    \vspace{-0.2cm}
    \begin{subfigure}{0.48\linewidth}
        \includegraphics[width=\textwidth]{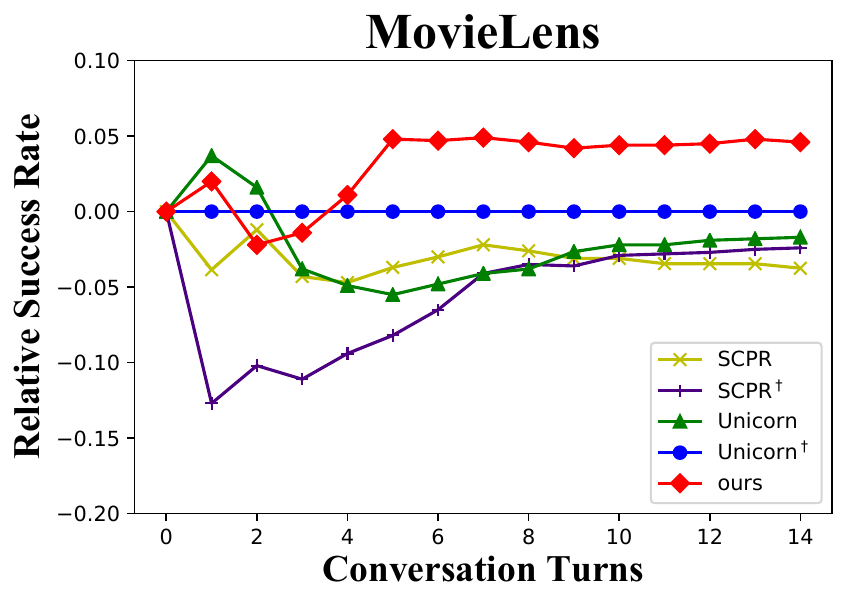}
    \end{subfigure}
     \vspace{-0.2cm}
    \caption{Comparisons at Different Conversation Turns.}
    \label{fig:overall}
    \vspace{-0.5cm}
\end{figure}
\subsubsection{Evaluation Metrics}
Following previous studies on MCR \cite{EAR,SCPR,Unicorn},  we utilize success rate (SR@$T$) \cite{CRM} to measure the cumulative ratio of successful recommendation with the maximum turn $T$, and average turn (AT) to evaluate the average number of turns. Besides, we adopt hDCG@($T$,$K$) \cite{Unicorn} to additionally evaluate the ranking performance of recommendations. For SR@$t$ and hDCG@($T$,$K$), the higher value indicates better performance. While the lower AT means the overall higher efficiency.


\subsection{Performance Comparison (RQ1)}
The comparison experimental results of the baseline models and our models are shown in \autoref{results}. We also intuitively present the performance comparison of success rate at each turn in \autoref{fig:overall}.
Relative success rate denotes the difference between each methods and the most competitive baseline $\text{UNICORN}^{\dagger}$, where the blue line of $\text{UNICORN}^{\dagger}$ is set to $y = 0$ in the figures. For clear observation, we only report the result of four competitive baselines and our model.
Based on the comparison in the table and figures, we can summarize our observations as follows:

\begin{table}[t]
    \caption{Results of the Ablation Study. }
    \vspace{-0.3cm}
    \label{tab:ablation_study}
    \small
    \centering
    \begin{tabular}{p{2.7cm}<{\centering}p{0.6cm}<{\centering}p{0.5cm}<{\centering}p{0.6cm}<{\centering}p{0.6cm}<{\centering}p{0.5cm}<{\centering}p{0.6cm}<{\centering}}
    \toprule
    \multirow{2}{*}{\bfseries Models }& \multicolumn{3}{c}{\textbf{Yelp}}& \multicolumn{3}{c}{\textbf{Amazon-Book}}\\
    &SR@15&AT&hDCG&SR@15&AT&hDCG\\
    \midrule
    Ours&0.482&11.87&0.160&0.545&10.83&0.223\\
    \midrule
    -w/o multi-interest&0.435&12.41&0.145&0.522&10.96&0.204\\
    -w/o global graph&0.463&12.31&0.150&0.516&11.03&0.198\\
    \midrule
    -binary questions&0.448&12.96&0.151&0.513&11.12&0.192\\
    -intersection set strategy&0.414&12.29&0.145&0.438&11.81&0.159\\
    -Sum-based strategy&0.467&11.94&0.152&0.529&11.01&0.217\\
    \bottomrule
    \end{tabular}
    \vspace{-0.5cm}
\end{table}

\begin{itemize}[leftmargin=*]
    \item \textbf{} Our framework outperforms all the comparison methods on four datasets. Compared with baselines, our method extends the form of questions to attribute type-based multiple choice formula, eliciting user's feedback of multi-acceptable items efficiently. Besides, the union set strategy can effectively avoid over-filtering items. Moreover, we extract multiple interests of the user from the accepted attribute instances by combining current preferences with historical interactions, instead of utilizing a mixed single state representation to decide the next action.
    
    \item \textbf{} Compared to the original version of SCPR and UNICORN, adapted $\text{SCPR}^{\dagger}$ and $\text{UNICORN}^{\dagger}$ achieve better performance, which indicates the effectiveness of above designs (multiple choice questions and union set strategy) for MIMCR. Nevertheless, our method still outperforms the adapted methods. We infer that the single user preference extracted by these baselines limits the ability to capture fine-grained user interests.
    \item \textbf{} Interestingly, we can find that original SCPR and UNICORN outperform adapted versions at the first few turns, but they fall quickly as the turn increases. Since original frameworks narrow the candidate item set following the intersection set strategy, and the acceptable items might not be filtered out when the number of accepted attribute instances is small, a smaller candidate item set can increase the probability of successful recommendation.
    As the number of conversations turn grows, the over-specific candidate item set over-filters out the acceptable items, which limits the subsequent improvement of these methods. On the contrary, our method achieves an outstanding performance in the latter stage of the conversation, where there are still comparatively generalized candidate items set and attributes space to avoid over-filtering. 
\end{itemize}

\subsection{Ablation Studies (RQ2)}

\begin{figure}[t]
    \centering
    \begin{subfigure}{0.48\linewidth}
        \includegraphics[width=\textwidth]{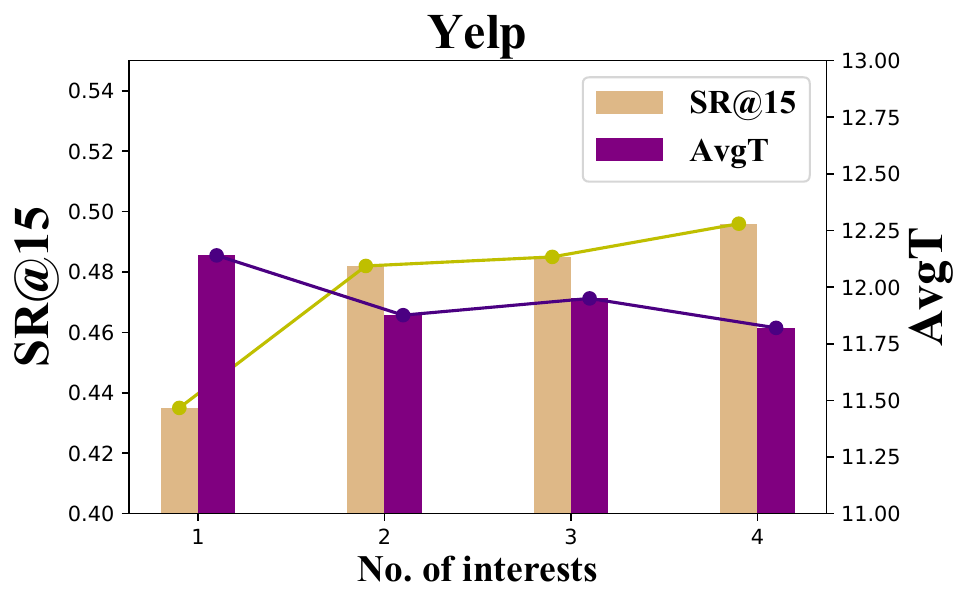}
    \end{subfigure}
    \vspace{-0.1cm}
    \begin{subfigure}{0.48\linewidth}
        \includegraphics[width=\textwidth]{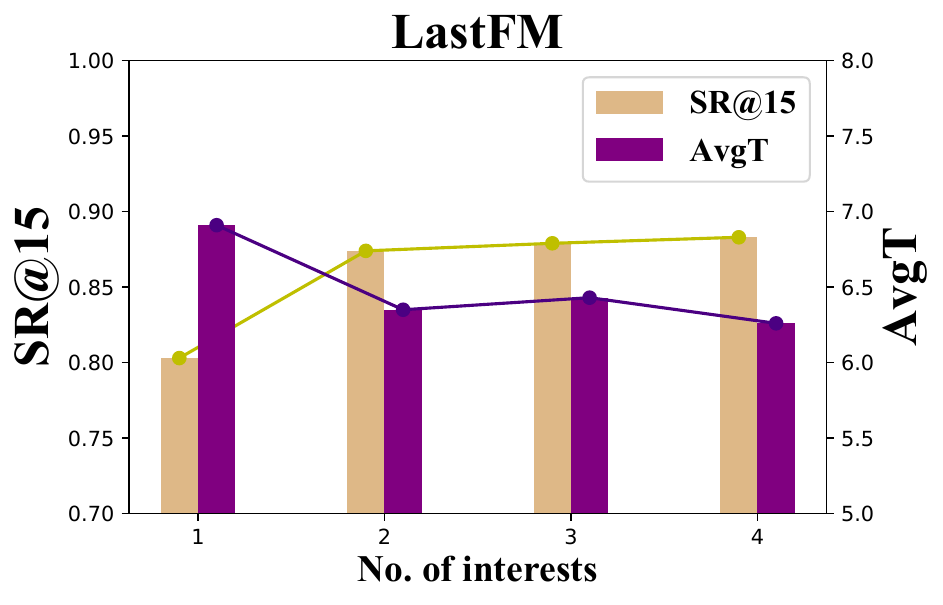}
    \end{subfigure}
    \vspace{-0.3cm}
    \caption{Performance comparisons w.r.t. $K_I$ with $N_v=2$.}
    \label{fig:interest2}

    \begin{subfigure}{0.48\linewidth}
        \includegraphics[width=\textwidth]{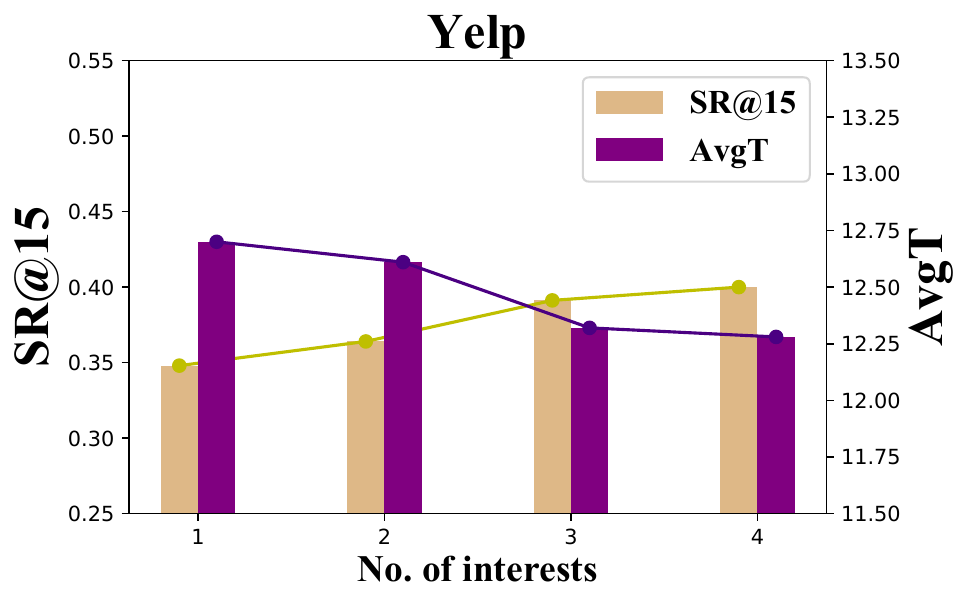}
    \end{subfigure}
    \vspace{-0.1cm}
    \begin{subfigure}{0.48\linewidth}
        \includegraphics[width=\textwidth]{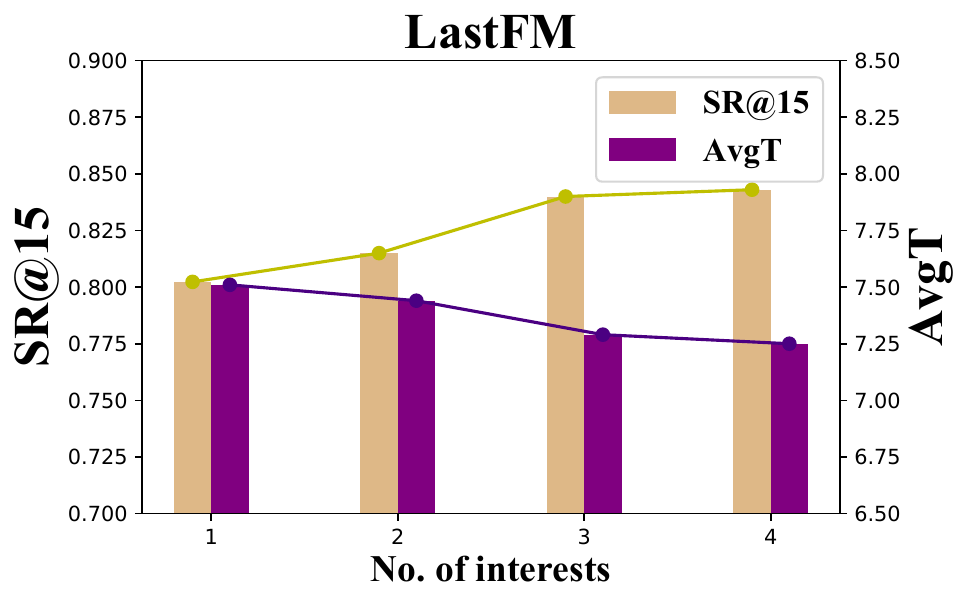}
    \end{subfigure}
    \vspace{-0.3cm}
    \caption{Performance comparisons w.r.t. $K_I$ with $N_v=3$.}
    \label{fig:interest3}
    \vspace{-0.4cm}
\end{figure}
In order to verify the effectiveness of some key designs, we conduct a series of ablation experiments on the Yelp and Amazon-Book datasets. The results are shown in Table~\ref{tab:ablation_study}.

\subsubsection{Impact of different modules}
Firstly, we evaluate the effectiveness of different modules, including Iterative Multi-interest Extractor and Global Graph Representation. Specifically, we remove these critical modules of MCMIPL to observe performance changes. As can be seen in Table~\ref{tab:ablation_study}, the model performance decreases significantly without the Iterative Multi-interest Extractor, which suggests that multi-interest representation is more appropriate for MIMCR, compared to the mixed single-interest representation.
Moreover, we can see that the removal of Global Graph Representation module also leads to poor performance, which indicates that the historical user representation is important for revealing latent user preferences and guiding the extraction of current multiple interests.


\subsubsection{Impact of different strategies.}
We conduct some experiments to study the effectiveness of strategies. Specifically, we retain the binary question type ("-binary questions"), traditional candidate item filtering strategy ("-intersection set strategy"), separately. Meanwhile, we utilize the Sum-based strategy to decide the attribute type involved in questions.
The binary question type version of our model performs worse than default setting, which demonstrates the efficiency of multiple choice question types for the conversational interaction. Besides, the intersection set strategy achieves inferior performance. It can be inferred that limitation of item selection strategy based on all accepted attribute instances will over-filter some user-acceptable items. While for the adjustment of sum-based strategy, the model still keeps competitive performance in all metrics for MIMCR, which indicates that this strategy can select suitable attribute types based on user interests.

\subsection{Hyper-parameter Sensitivity Analysis (RQ3)}
\subsubsection{Impact of Interests Number}
Since interest number $K_I$ is closely related to maximum number $N_v$ of acceptable items. We explore the hyper-parameter $K_I$ in the case of the maximum number $N_v$ of acceptable items is 2 and 3 respectively. As we can see from Figure~\ref{fig:interest2} and Figure~\ref{fig:interest3}, with the increase of interest number $K_I$, the performance of our methods improves. In addition, when the interest number $ K_I $ exceeds the maximum number of acceptable items, the performance will hardly improve, which indicates that some interests may exist redundancy and point to the same user preferences.

\subsubsection{Impact of Asked Attribute Instances Number}
\begin{figure}[t]
    \centering
    \begin{subfigure}{0.48\linewidth}
        \includegraphics[width=\textwidth]{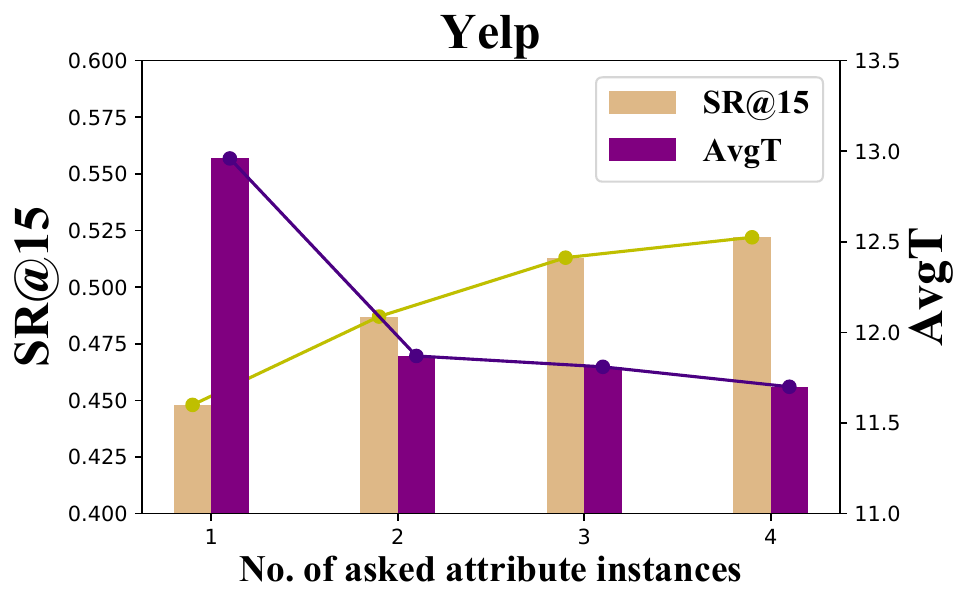}
    \end{subfigure}
    \vspace{-0.1cm}
    \begin{subfigure}{0.48\linewidth}
        \includegraphics[width=\textwidth]{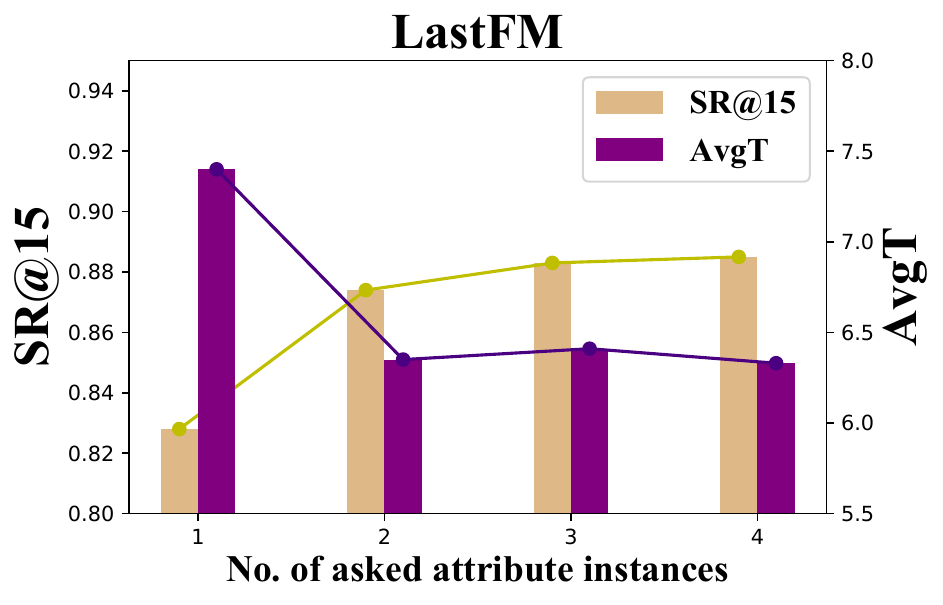}
    \end{subfigure}
    \vspace{-0.3cm}
    \caption{Performance comparisons w.r.t. $K_a$.}
    \label{fig:parameter-ask}
    \vspace{-0.4cm}
\end{figure}

When asking users questions, the attribute instances number $K_a$ included in a question affects model performance. 
As can be seen in Figure~\ref{fig:parameter-ask}, the performance improves as the value of $K_a$ increases, which indicates that the larger number of asked attribute instances in a turn, the more information the CRS obtains. However, if the value of $K_a$ is too large, performance improvement is limited. That also indicates the most of extra attribute instances are invalid.

\begin{figure}[t]
    \centering
    \includegraphics[width=0.46\textwidth]{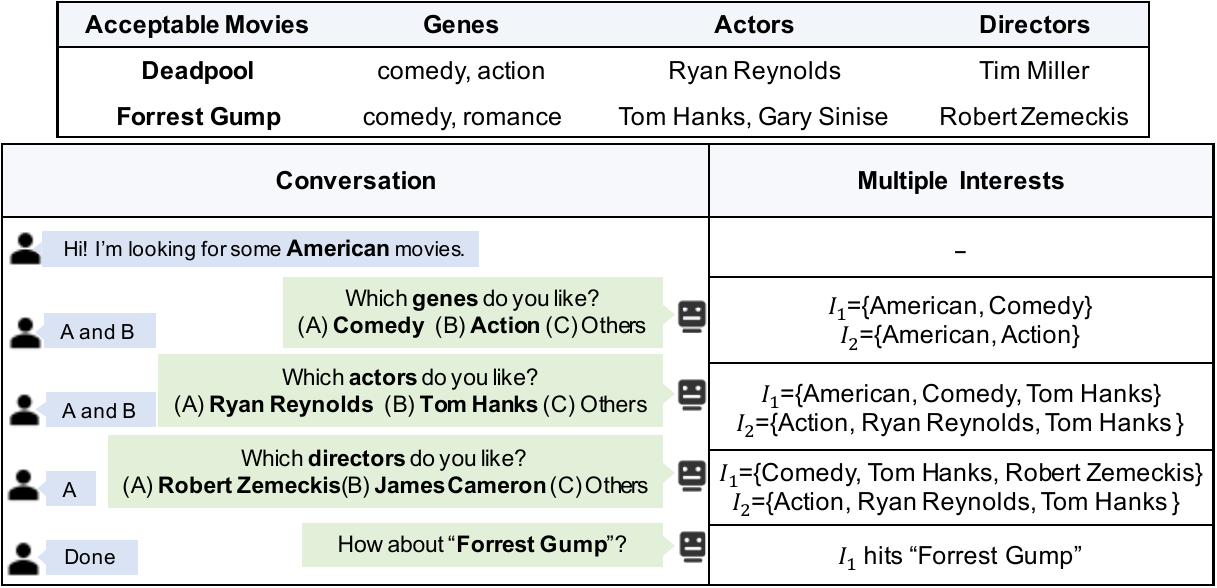}
    \vspace{-0.3cm}
    \caption{A conversation generated by our framework. $I_1$ and $I_2$ denote two interests of the user, respectively.
    }
    \label{fig:case_study)}
    \vspace{-0.4cm}
\end{figure}

\subsection{Case Study (RQ4)}

To show the process of extracting the user's multiple interests, we present a conversation case generated by our framework from MovieLens dataset in \autoref{fig:case_study)}. We only show the attribute types and instances that are relevant to the questions.
For each interest, we present attribute instances with high contribution rate, where the sum of their attention scores $\geq 0.8$.
As can be seen, based on user's feedback of each turn as well as historical global information, our model extracts multiple interests in different attribute instances combinations.
Finally, our method makes a successful recommendation based on one of the interest representations that perfectly matches user's preference.

\section{Conclusion}
In this work, we define a more realistic CRS scenario named MIMCR, in which the user may accepts one of multiple potential items instead of single target item in a conversation. Based on the scenario, we propose a novel framework MCMIPL, which generates multiple choice questions to collect user preferences, and utilizes union set strategy to select candidate items. In addition, we propose a MIPL module to exact multi-interest of the user to decide the next action. Extensive experimental results on four datasets demonstrate the superiority of our method in the proposed scenario.





\begin{acks}
The work is partially supported by the National Nature Science Foundation of China (No. 61976160, 61976158, 61906137), Shanghai Science and Technology Plan Project (No. 21DZ1204800) and Technology research plan project of Ministry of Public and Security (Grant No. 2020JSYJD01).
\end{acks}

\bibliographystyle{ACM-Reference-Format}
\bibliography{sample-base}

\end{document}